\documentclass[%
 reprint,
 amsmath,amssymb,
 aps,
 prb,
]{revtex4-1}

\usepackage{graphicx}
\usepackage{subfigure}
\usepackage{dcolumn}
\usepackage{bm}

\begin{document}

\preprint{APS/123-QED}

\title{Influences of weak disorder on dynamical quantum phase transitions of anisotropic XY chain}

\author{Kaiyuan Cao}
\affiliation{Department of Physics and Institute of Theoretical Physics, Nanjing Normal University, Nanjing 210023, People¡¯s Republic of China}

\author{Wenwen Li}
\affiliation{Department of Physics and Institute of Theoretical Physics, Nanjing Normal University, Nanjing 210023, People¡¯s Republic of China}

\author{Ming Zhong}
\affiliation{Department of Physics and Institute of Theoretical Physics, Nanjing Normal University, Nanjing 210023, People¡¯s Republic of China}

\author{Peiqing Tong}
\email{pqtong@njnu.edu.cn}
\affiliation{Department of Physics and Institute of Theoretical Physics, Nanjing Normal University, Nanjing 210023, P. R. China and Jiangsu Key Laboratory for Numerical Simulation of Large Scale Complex Systems, Nanjing Normal University, Nanjing 210023, P. R. China}

\date{\today}

\begin{abstract}
  In this paper, the effects of disorder on the dynamical quantum phase transitions (DQPTs) in the transverse-field anisotropic XY chain are studied by numerically calculating the Loschmidt echo after quench. We obtain the formula for calculating the Loschmidt echo of the inhomogeneous system in real space. By comparing the results with that of the homogeneous chain, we find that when the quench crosses the Ising transition, the small disorder will cause a new critical point. As the disorder increases, more critical points of the DQPTs will occur, constituting a critical region. In the quench across the anisotropic transition, the disorder will cause a critical region near the critical point, and the width of the critical region increases by the disordered strength. In the case of quench passing through two critical lines, the small disorder leads to the system to have three additional critical points. When the quench is in the ferromagnetic phase, the large disorder causes the two critical points of the homogeneous case to become a critical region. And for the quench in the paramagnetic phase, the DQPTs will disappear for large disorder.
\end{abstract}

\maketitle


\section{\label{sec:level1} Introduction}

Recently, the nonequilibrium dynamics have attracted a lot of interest in quantum many-body systems \cite{r1,r2,r3,r4,r5,r6}. One of the interesting topics is to study the non-analytical behavior of physical quantities during the dynamical evolution with time. Such singularity phenomenon is suggested as dynamical quantum phase transition (DQPT) \cite{r7}. By considering the formal similarity with the canonical partition function in the equilibrium system, the overlap amplitude (Loschmidt amplitude) of the initial state with the evolved time state is regraded as the dynamical free energy in the quantum system. It is found that the Loschmidt amplitude ${\cal G}(t)$ and the return probability Loschmidt echo ${\cal L}(t)$ become non-analytic when the magnitic field is quenched through the critical point in the transverse-field quantum Ising chain \cite{r7}. The similar singularity is also found in many systems, including the extended transverse-field Ising chain \cite{r8}, XY model \cite{r9,r10,r11,r011}, XXZ chain \cite{r12,r13,r14,r15}, kicked quantum Ising chain \cite{r16}, the Kitaev honeycomb model \cite{r016}, open Dicke model \cite{r17,r18} and two-banded topological systems \cite{r19,r20,r21,r22,r23,r024}, etc. The results reveal that nonanalytic dynamics is a generic feature of sudden quenches across quantum critical points. However, a substantial number of examples demonstrate that the DQPTs may occur without crossing quantum phase transitions (QPTs) \cite{r10,r011}. In addition, the DQPTs have been observed in ultra-cold atom and quantum simulators experiments \cite{r24,r25,r26,r27,r28}.

While most theoretical works on DQPTs focus on the quantum systems with phase transitions caused by broken symmetries or accompanied by the close of energy gap, the DQPTs driven by disorder are rarely discussed. This is a particularly important on account of disorder has a drastic effect on the equilibrium phase transition (EPT) and quantum phase transition (QPT). A large number of first-order phase transitions are smeared and developed into a smooth crossover in the presence of disorder \cite{r29,r30,r31}. For continuous transitions, the Harris criterion can be used to predict the stability of a transition against weak disorder by the critical exponents \cite{r32}. On the other hand, it is well known that the disorder can cause the extended state of electrons to Anderson's localized state \cite{r33}. Although there are a little works on DQPTs in inhomogeneous systems \cite{r34,r35}, it is still unclear to what extent the disorder influences the DQPTs on a general level.

In this paper, we study the influences of disorder on the DQPTs in the transverse-field anisotropic XY chain. In our work, five types of quench protocols that can cause the DQPTs in the homogeneous chain are used as references, including quench across the Ising transition or anisotropic transition, quench across two kinds of quantum transitions and quench inside the same phase (the ferromagnetic phase or the paramagnetic phase). We will discuss what effects on the DQPTs with the weak disordered exchange interactions $\{J_{n}\}$ in the system. The paper is organized as following: In Sec. \uppercase\expandafter{\romannumeral2}, we introduce the model and the method to calculate the Loschmidt echo in the disorder systems. The numerical results are given in Sec. \uppercase\expandafter{\romannumeral3}. The Sec.\uppercase\expandafter{\romannumeral4} is a short summary.

\section{Model and Loschmidt echo}

The general Hamiltonian of the transverse-field anisotropic XY chain is expressed as
\begin{equation}\label{eq:1}
  H= -\frac{1}{2}\sum_{n=1}^{N}\{\frac{J_{n}}{2}[(1+\gamma)\sigma_{n}^{x}\sigma_{n+1}^{x}+
  (1-\gamma)\sigma_{n}^{y}\sigma_{n+1}^{y}]+h\sigma_{n}^{z}\},
\end{equation}
where $\gamma$ is the anisotropic coefficient, $h$ represents the transverse field, $J_{n}$s are exchange couplings between the nearest-neighbour spins, respectively. We take $J_{n}=J+\Delta J_{n}$, where $\Delta J_{n}$ is an independent random number distributed uniformly in the interval $[-w/2,w/2]$ with $w$ denoting the strength of disorder. For convenience we take $J=1$ without loss of generality. By the way, the Hamiltonian (\ref{eq:1}) describes the homogeneous system in the case $w=0$.

In a quantum quench, the system is prepared in the ground state $|\psi_{0}\rangle$ of the initial Hamiltonian $H=H(\gamma_{0},h_{0})$. Then, at time $t=0$, the parameters $(\gamma_{0},h_{0})$ are suddenly changed to $(\gamma_{1},h_{1})$ of the post-quench Hamiltonian $\tilde{H}=H(\gamma_{1},h_{1})$. The Loschmidt amplitude after the quench is given by
\begin{equation}\label{eq:2}
  {\cal G}(t)=\langle\psi_{0}|e^{iHt}e^{-i\tilde{H}t}|\psi_{0}\rangle,
\end{equation}
with the associated Loschmidt echo ${\cal L}(t)=|{\cal G}(t)|^{2}$.

The Hamiltonian (\ref{eq:1}) can be mapped to a quadratic spinless fermion model by the Jordan-Wigner transformation \cite{r39}. Then both the pre-quench Hamiltonian $H$ and the post-quench Hamiltonian $\tilde{H}$ can be reduced to the diagonal forms by the general Bogoliubov transformation in the real space \cite{r40}
\begin{equation}\label{eq:3}
  H= \sum_{k}\Lambda_{k}(\eta^{\dag}_{k}\eta_{k}-\frac{1}{2}),
\end{equation}
and
\begin{equation}\label{eq:4}
  \tilde{H}= \sum_{k}\tilde{\Lambda}_{k}(\tilde{\eta}^{\dag}_{k}\tilde{\eta}_{k}-\frac{1}{2}),
\end{equation}
respectively.

The Fermi operator $\eta_{k}$ of the pre-quench Hamiltonian $H$ can be expressed by the operators $\tilde{\eta}^{\dag}_{i}$ and $\tilde{\eta}_{i}$ of the post-quench Hamiltonian $\tilde{H}$ as
\begin{equation}\label{eq:5}
  \eta_{k}=\sum_{i}(g_{ki}\tilde{\eta}^{\dag}_{i}+h_{ki}\tilde{\eta}_{i}).
\end{equation}
Suppose $|\tilde{\psi}_{0}\rangle$ is the ground state of the post-quench Hamiltonian $\tilde{H}$, the initial state $|\psi_{0}\rangle$ can be written as \cite{r41}
\begin{equation}\label{eq:6}
  |\psi_{0}\rangle=\frac{1}{{\cal N}}\exp{(\frac{1}{2}\sum_{i,j}\tilde{\eta}^{\dag}_{i}G_{ij}\tilde{\eta}^{\dag}_{j})}
  |\tilde{\psi}_{0}\rangle,
\end{equation}
where the antisymmetry matrix $G$ satisfies the equations
\begin{equation}\label{eq:7}
  \sum_{i}g_{ki}G_{ij}+h_{kj}=0, \quad j, k= 1,\cdots,N.
\end{equation}

By substituting the Eq.~(\ref{eq:6}) into Eq.~(\ref{eq:2}), the Loschmidt amplitude ${\cal G}(t)$ will be given as
\begin{equation}\label{eq:8}
  {\cal G}(t)=\frac{e^{i(E_{0}-\tilde{E}_{0})t}}{{\cal N}}\prod_{j,k>j}[1+e^{it(\tilde{\Lambda}_{j}+\tilde{\Lambda}_{k})}G_{jk}^{2}],
\end{equation}
where ${\cal N}^{2}=\Pi_{j,k>j}(1+G_{jk}^{2})$ is the normalization coefficient. Then the Loschmidt echo is
\begin{equation}\label{eq:9}
  \begin{split}
    {\cal L}(t) & ={\cal G}^{*}(t){\cal G}(t) \\
      & =\prod_{j,k>j}[1-\frac{4G_{jk}^{2}}{(1+G_{jk}^{2})}\sin^{2}{(\frac{\tilde{
      \Lambda}_{j}+\tilde{\Lambda}_{k}}{2}t)}].
  \end{split}
\end{equation}

The rate function $\lambda(t)$, served like the free energy function in equilibrium phase transition, is defined to reflect the DQPTs more directly. At the critical time, the rate function of the Loschmidt echo exhibits a nonanalytic kink \cite{r38}. According to the definition,
\begin{equation}\label{eq:10}
 \begin{split}
   \lambda(t) & =-\lim_{N\rightarrow\infty}\frac{1}{N}\ln{[{\cal L}(t)]}  \\
     & = -\lim_{N\rightarrow\infty}\frac{1}{N}\sum_{j,k>j}\ln{[1-\frac{4G_{jk}^{2}}{(1+G_{jk}^{2})^{2}}\sin^{2}(\frac{\tilde{
      \Lambda}_{j}+\tilde{\Lambda}_{k}}{2}t)]}.
 \end{split}
\end{equation}

On the other hand, Heyl et al. \cite{r7} have pointed out that the critical times on the rate function correspond to the Fisher zeros of Loschmidt amplitude ${\cal G}(t)$ on the imaginary axis in the complex time plane. From Eq.~(\ref{eq:8}), we can obtain the Fisher zeros of the Loschmidt amplitude ${\cal G}(t)$ in the complex time plane expressed by a family of points labeled by a number $n\in\mathbb{Z}$
\begin{equation}\label{eq:11}
  it_{n}= \frac{1}{\tilde{\Lambda}_{j}+\tilde{\Lambda}_{k}}[\ln{G_{jk}^{2}}+i(2\pi+1)n],
\end{equation}
where $i$ denotes imaginary unit. Those Fisher zeros on the imaginary axis correspond to the critical times of the DQPTs. So with $G^{2}_{jk}=1$, the critical times are
\begin{equation}\label{eq:12}
  t_{n}^{*}=t^{*}_{0}(2n+1),
\end{equation}
where $t^{*}_{0}=\pi/(\tilde{\Lambda}_{j}+\tilde{\Lambda}_{k})$. Therefore, we only need to get the first critical time, and other critical times can be calculated by Eq.~(\ref{eq:12}).

\begin{figure}
  \centering
  \includegraphics[width=1.0\linewidth]{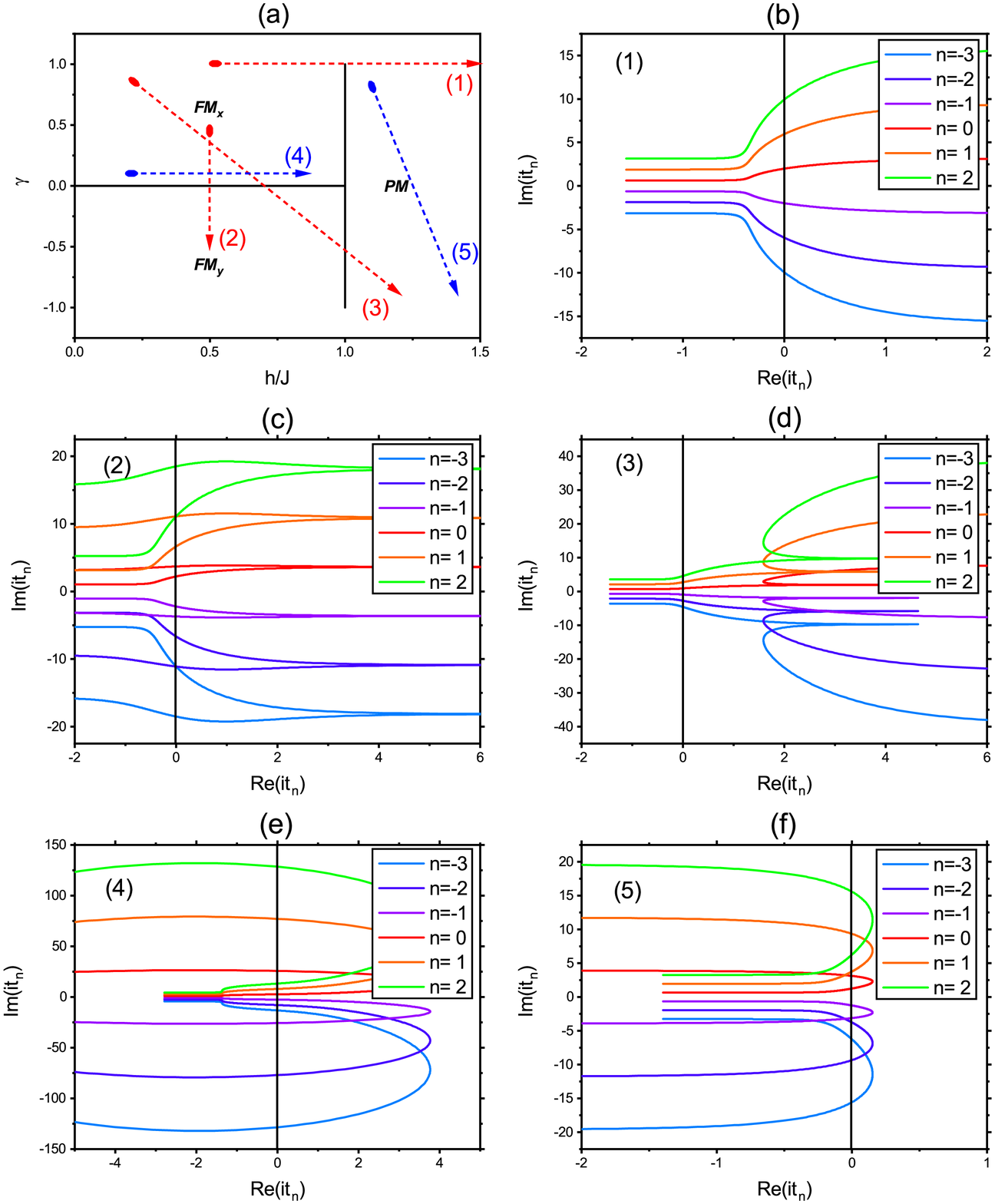}\\
  \caption{ (a) The phase diagram of the homogeneous XY chain, in which five typical quench paths (1)-(5) are marked. (b)-(f) are the corresponding diagrams of Fisher zeros in complex time plane. The critical times for different $n$ are denoted by the intersections of the Fisher zeros lines and the imaginary axis. }\label{fig:01}
\end{figure}

 For the convenience to the reader, the DQPTs of the homogeneous chain as a reference will be briefly introduced in the following. The Loschmidt amplitude ${\cal G}(t)$ of the homogeneous system has already been given analytically in momentum space \cite{r42}
\begin{equation}\label{eq:013}
  {\cal G}(t)= e^{i(E_{0}+\tilde{E}_{0})t}\prod_{k>0}(\cos^{2}{\alpha_{k}}+\sin^{2}{\alpha_{k}}e^{-2i\tilde{
  \Lambda}_{k}t}).
\end{equation}
The corresponding Fisher's zeros of Eq.~(\ref{eq:013}) are located along the lines $n\in\mathbb{Z}$:
\begin{equation}\label{eq:14}
  it_{n}=\frac{1}{2\tilde{\Lambda}_{k}}[\ln{\tan^{2}{\alpha_{k}}}+i(2\pi+1)n].
\end{equation}
From $\tan^{2}{\alpha_{k^{*}}}=1$, the $k^{*}$ corresponding to the critical time satisfies the following quadratic equation
\begin{equation}\label{eq:15}
  (1-\gamma_{0}\gamma_{1})\cos^{2}{k^{*}}+(h_{0}+h_{1})\cos{k^{*}}+(h_{0}h_{1}+\gamma_{0}\gamma_{1}
  )=0,
\end{equation}
Obviously, the solutions of Eq.~(\ref{eq:15}) are determined by the values of $h_{0}$, $h_{1}$ and $\gamma_{0}\gamma_{1}$. By analyzing the solutions, we can judge whether the DQPTs exist for a exact quench process. In Fig.~\ref{fig:01} (a), we show five quench paths which can cause the DQPTs in the homogeneous XY chain. And the corresponding Fisher zeros are shown in Fig.~\ref{fig:01} (b)-(f). In the diagram of Fisher zeros, each curve corresponds to the Fisher zeros of $n$, so that we call it Fisher zeros line for convenience. The intersections of the Fisher zeros line and the imaginary axis in the complex time plane denote the critical times of the DQPTs. If there are two intersections with imaginary axis on each Fisher zeros line, it means that two kinds of the DQPTs exist in this quench process [see Fig.~\ref{fig:01} (b), (e), (f)].

\section{Numerical Results}

In this section, we study the influences of disorder on the DQPTs for five typical quench protocols without losing generality. The number of spins is taken as $N=1000$, and the larger $N$ gives the similar results. We diagonalize the pre-quench Hamiltonian and the post-quench Hamiltonian by numerical method, and construct the matrix $G$ with the help of Bogoliubov transformation in real space \cite{r41}.

\subsection{quench across the Ising transition}

\begin{figure}[b]
  \centering
  \includegraphics[width=1.0\linewidth]{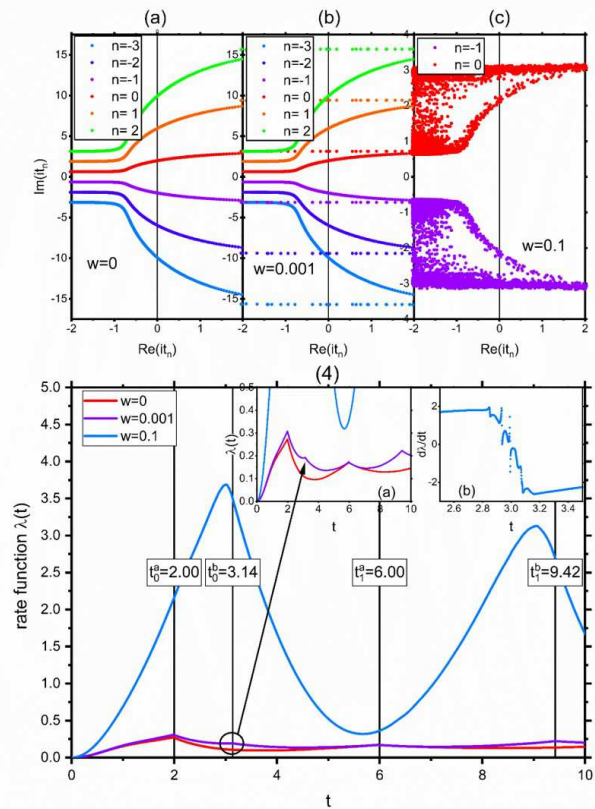}\\
  \caption{(a)-(c) The Fisher zeros of the homogeneous chain ($w=0$) and two disorder chains with $w=0.001$ and $0.1$ for a quench across the Ising transition from $(\gamma_{0}=1.0,h_{0}=0.5)$ to $(\gamma_{1}=1.0,h_{1}=1.5)$. (d) The rate functions of Loschmidt echo for different disorder strengths $w$. The insert graph (i) is the enlarged graph of the rate functions in order to see the cases of $w=0$ and $0.001$. The insert graph (ii) is the first order derivative of rate function $\texttt{d}\lambda/\texttt{d}t$ for the disorder chain ($w=0.1$).}\label{fig:02}
\end{figure}

\begin{figure}[t]
  \centering
  \includegraphics[width=1.0\linewidth]{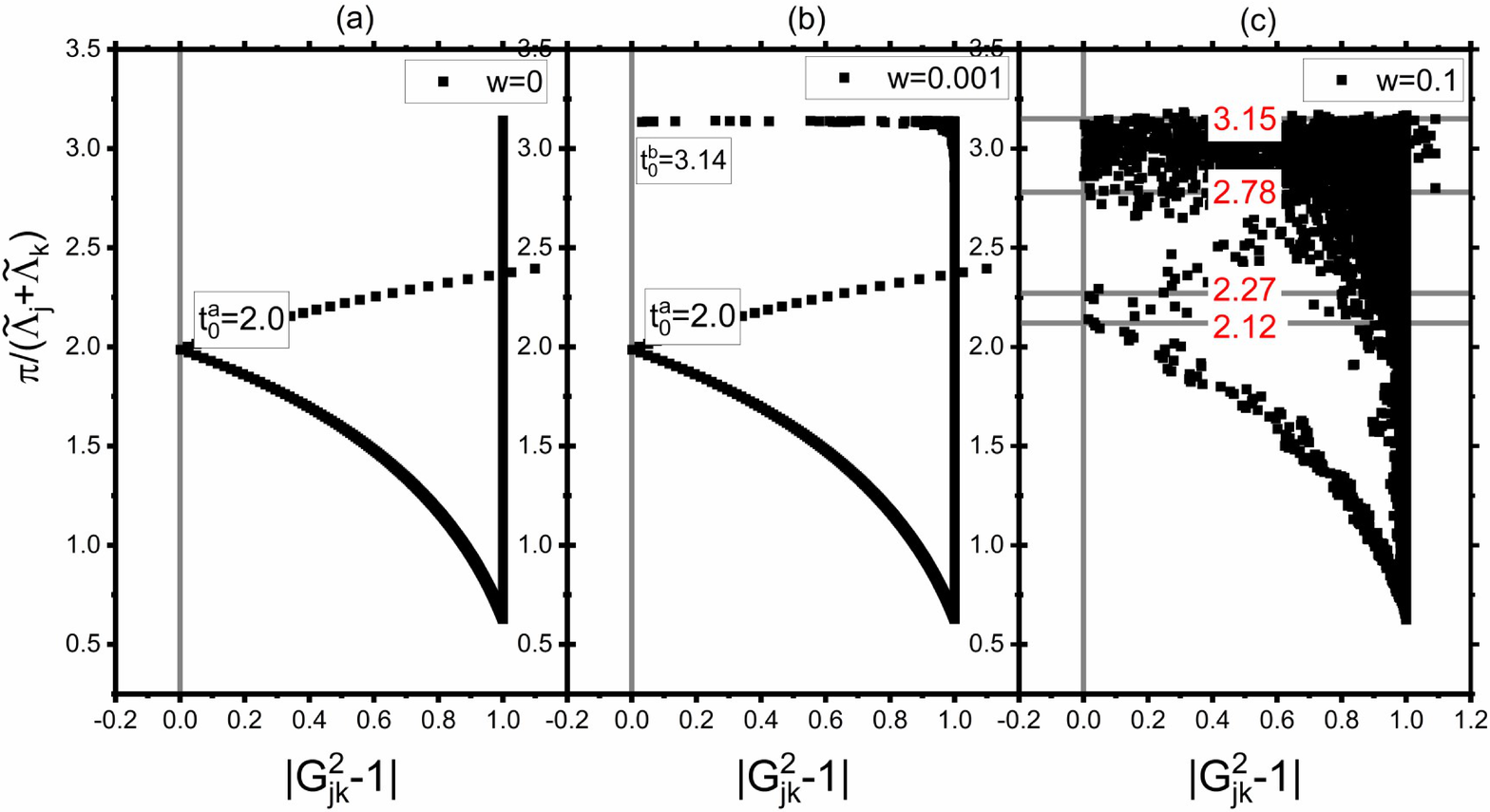}\\
  \caption{The relationships between the quantity $\pi/(\tilde{\Lambda}_{j}+\tilde{\Lambda}_{k})$ and $|G^{2}_{jk}-1|$ of the chains with disorder strengths $w=0, 0.001$ and $0.1$, for the quench across Ising transition. The intersections of the curve and the zero line $|G^{2}_{jk}-1|=0$ denote the critical times of the system. The numbers on the grey lines are values of the critical times. }\label{fig:03}
\end{figure}

First of all, let us consider the case of quench across the Ising transition. The Fig.~\ref{fig:02} shows the Fisher zeros and corresponding rate functions in the homogeneous chain ($w=0$) and two disorder chains with $w=0.001$ and $0.1$, respectively.  In Fig.~\ref{fig:02} (a), we can see that the Fisher zeros of the homogeneous chain, obtained by numerical method in real space, show great agreements with that by analytical method in momentum space [see Fig.~\ref{fig:01} (b)]. There is an intersection of every Fisher zeros line and the imaginary axis in complex time plane, which corresponds the critical times $t^{a}_{0}, 3t^{a}_{0}, 5t^{a}_{0}, \cdots$, with $t^{a}_{0}\approx2.0$. From the rate function in Fig.~\ref{fig:02} (d), the sharp structures appear at these critical times. In the presence of small disorder, a straight line appears on the outside of every Fisher zeros line [see Fig.~\ref{fig:02} (b) for  $w=0.001$], so that there are new additional intersections corresponding to the critical time $t^{b}_{n}=t^{b}_{0}(2n+1)$, with $t^{b}_{0}\approx3.14$. Meanwhile, we can see a new born kink at the critical time $t^{b}_{0}$ in the insert graph (i) of Fig.~\ref{fig:02} (d). The results reveal that the small disorder will cause a new critical point. As the disorder strength increasing, more and more Fisher zeros will appear on the imaginary axis to constitute some continuous regions, which can be seen in Fig.~\ref{fig:02} (c) for $w=0.1$.  At the same time, the corresponding rate function [see the blue line in Fig.~\ref{fig:02} (d)] becomes smooth crossover near the critical time $t^{b}_{0}$. By studying the derivative of the rate function, we find that the derivative $d\lambda/dt$ has some oscillations and multiple mutation points in the critical region [see the insert graph (ii) of Fig.~\ref{fig:02} (d)]. This  demonstrates that the large disorder causes the Loschmidt echo function to exhibit non-analytic behavior in some time intervals.

In order to see the critical times more clearly, we analyze the relationship between the critical times and the transformation matrix $G$. In Fig.~\ref{fig:03}, we show the quantity $\pi/(\tilde{\Lambda}_{j}+\tilde{\Lambda}_{k})$ varying with the matrix elements correlated quantity $|G^{2}_{jk}-1|$ . According to Eqs.~(\ref{eq:11}) and (\ref{eq:12}), we know that the DQPTs occur where the elements $G_{jk}$ of antisymmetry matrix $G$ are equal to $\pm1$, so that the values $\pi/(\tilde{\Lambda}_{j}+\tilde{\Lambda}_{k})$ are the critical times when $|G^{2}_{jk}-1|=0$. The homogeneous system $w=0$ only has one the critical time $t^{a}_{0}\approx2.0$. While for the disorder system $w=0.001$, there are two critical times $t^{a}_{0}\approx2.0$ and $t^{b}_{0}\approx3.14$, respectively. When the disorder strength is increased to $w=0.1$, we can see two critical time regions at $t^{a}_{0}\in[2.12,2.27]$ and $t^{b}_{0}\in[2.79,3.15]$. This clearly confirms the above conclusions.

\subsection{quench across the anisotropic transition}

\begin{figure}[t]
  \centering
  \includegraphics[width=0.98\linewidth]{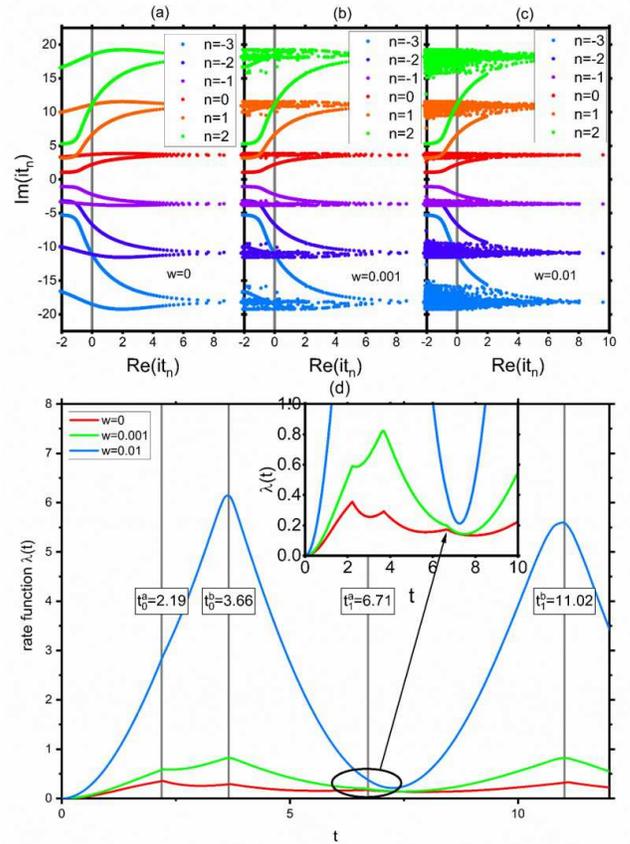}\\
  \caption{(a)-(c) The Fisher zeros of the homogeneous ($w=0$) and two disorder chains with $w=0.001$ and $0.01$ for the quench across the anisotropic transition from $(\gamma_{0}=0.5, h_{0}=0.5)$ to $(\gamma_{1}=-0.5, h_{1}=0.5)$. (d) The rate functions for different disorder strengths. The insert graph is the enlarged graph of the rate functions.}\label{fig:04}
\end{figure}

Secondly, the case of quench across the anisotropic transition is concerned. The Fisher zeros and corresponding rate functions are shown in Fig.~\ref{fig:04}. In the homogeneous chain, there are already two intersections on the each Fisher zeros line with the imaginary axis. And the corresponding critical times are $t^{a}_{0}\approx2.19$ and $t^{b}_{0}\approx3.66$, respectively. For the disorder chain, even with very small disorder strength, it can cause many intersections to form a continuous region. Such as the case of $w=0.001$, we can see these Fisher zeros region near the critical time $t^{b}_{n}$ on the imaginary axis [see Fig.~\ref{fig:04} (b)]. The width of critical regions increases with the increasing of the disorder strength [see Fig.~\ref{fig:04} (c)]. Unlike that in the case of quench passing through the Ising transition, the disorder does not introduce new critical points of DQPTs, but expands the critical time into a critical region.

\subsection{quench across the Ising transition and anisotropic transition}

\begin{figure}
  \centering
  \includegraphics[width=1.0\linewidth]{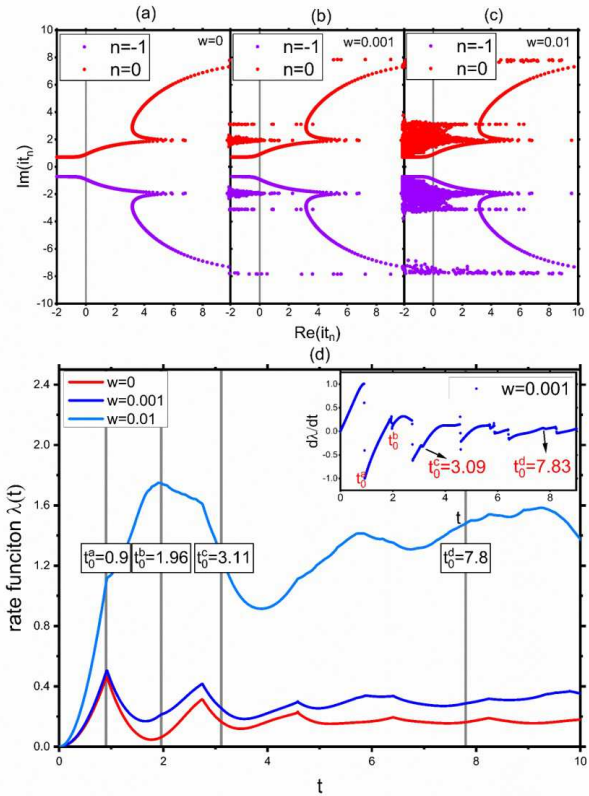}\\
  \caption{(a)-(c) The Fisher zeros of the homogeneous chain ($w=0$) and two disorder chains $w=0.001$ and $0.01$ for the quench across the Ising transition and anisotropic transition from $(\gamma_{0}=0.89, h_{0}=0.2)$ to $(\gamma_{1}=-0.89, h_{1}=1.2)$.  (d) The rate functions for different disorder strength. The insert graph is the derivative of rate function $\texttt{d}\lambda/\texttt{d}t$ of $w=0.001$. A small mutation is easily overlooked at $t=7.8$.  }\label{fig:05}
\end{figure}

\begin{figure}[t]
  \centering
  \includegraphics[width=0.99\linewidth]{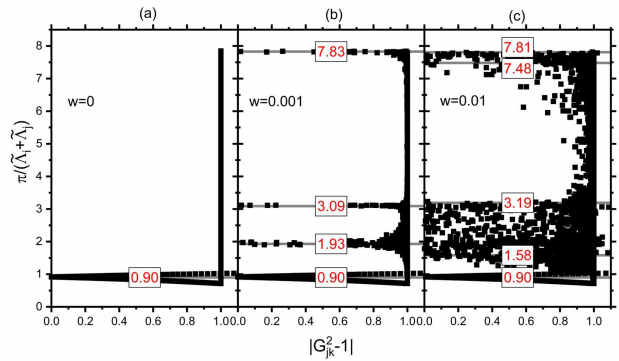}\\
  \caption{The $\frac{\pi}{\tilde{\Lambda}_{j}+\tilde{\Lambda}_{k}}$ evolving by $|G^{2}_{jk}-1|$ of the chains with $w=0, 0.001$ and $0.01$, for the quench paths across two kinds of quantum transitions.}\label{fig:06}
\end{figure}

In this section, we consider the case of quench across the Ising and the anisotropic transitions. The Fisher zeros and the corresponding rate functions of the homogeneous chain ($w=0$) and two disorder chains with $w=0.001$ and $0.01$ are shown in Fig.~\ref{fig:05}, respectively. In the homogeneous chain, one intersection can  be seen at each Fisher zeros line. The corresponding critical time is $t^{a}_{0}\approx0.90$. For the disorder chain, the weak disorder will cause three additional critical times $t^{b}_{0}\approx1.93, t^{c}_{0}\approx3.09$ and $t^{d}_{0}\approx7.83$. From Fig.~\ref{fig:05} (b), it is clearly seen that there are four intersections on each Fisher zeros line with the imaginary axis for $w=0.001$. Although it is difficult to see the critical points from the rate function, the derivatives of rate function $\texttt{d}\lambda/\texttt{d}t$ are discontinuous at the corresponding critical times [see the insert graph of Fig.~\ref{fig:05} (d)]. This indicates that the small disorder will cause the system to generate additional three critical points of the DQPTs. For the disorder strength increaseing further, more and more Fisher zeros will appear on the imaginary axis near the three new born critical times, and the critical points form the critical time regions. For example, in the case of $w=0.01$,  the middle two Fisher zeros regions are very close, and they merge to one region [see Fig.~\ref{fig:05} (c)]. Similarly, a clearer picture can be seen from the relationship between $\pi/(\tilde{\Lambda}_{j}+\tilde{\Lambda}_{k})$ and $|G^{2}_{jk}-1|$ [Fig.~\ref{fig:06}]. In the presence of small disorder $w=0.001$, there are indeed three additional critical times $1.93, 3.09$ and $7.83$. And the critical times are in good agreements with the results from the derivative $\texttt{d}\lambda/\texttt{d}t$. For the disorder chain $w=0.01$, we can see that the middle two critical times merge into one critical region $[7.48,7.81]$. In short summary, for case of the quench across two quantum transitions, the small disorder causes three additional critical points in the system. As the disorder increaseing, these critical times will constitute critical regions.

\begin{figure}[t]
  \centering
  \includegraphics[width=0.99\linewidth]{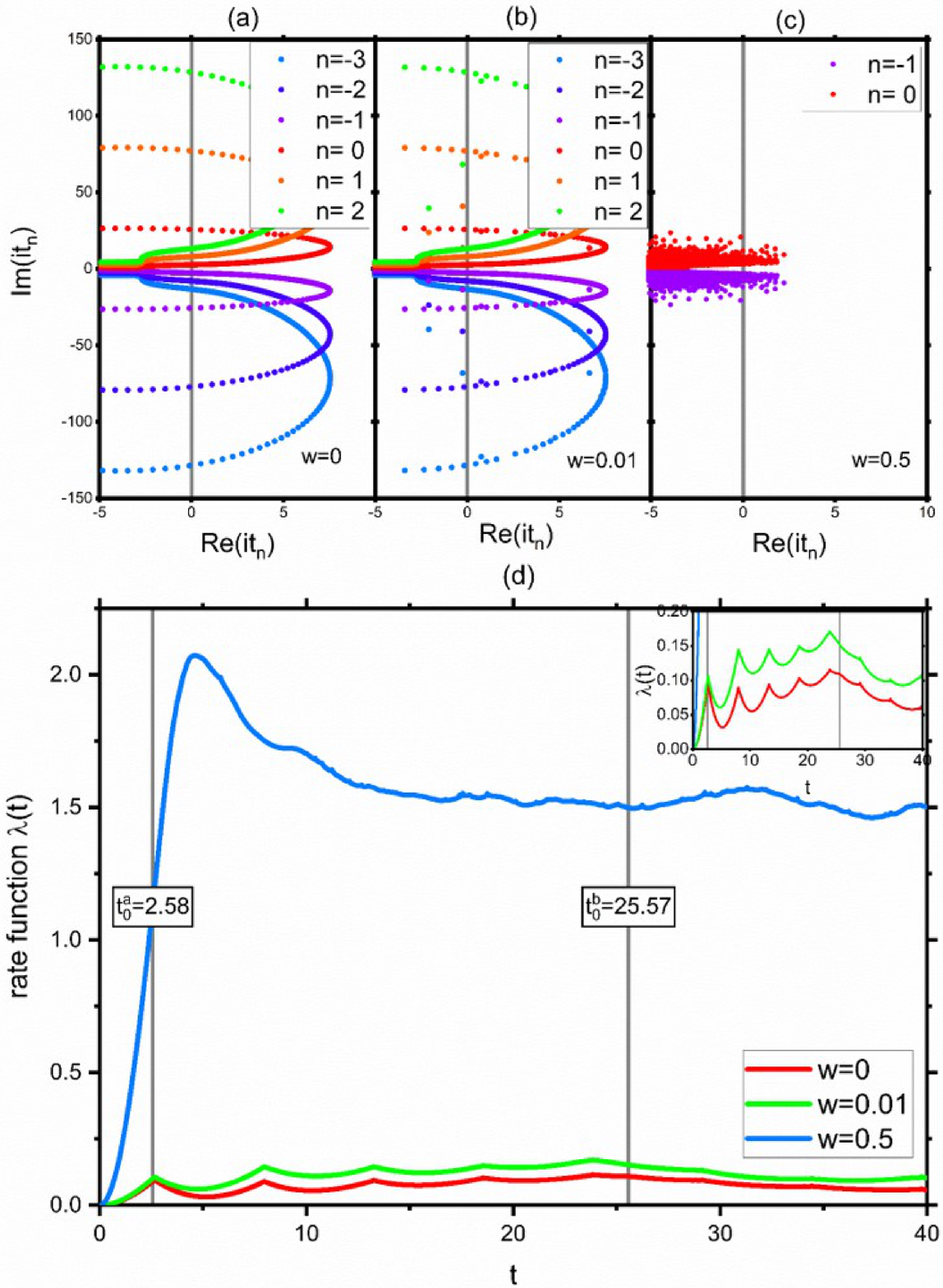}\\
  \caption{(a)-(b) The Fisher zeros of the homogeneous chain ($w=0$) and two disorder chains with $w=0.01$ and $0.5$ for the quench within the ferromagnetic phase from $(\gamma_{0}=0.1, h_{0}=0.2)$ to $(\gamma_{1}=0.1, h_{1}=0.8)$. (d) The corresponding rate functions for different disorder strengths. The insert graph is the enlarged graph of the rate functions.}\label{fig:07}
\end{figure}

\subsection{quench in the ferromagnetic phase.}

According to the Eq.~(\ref{eq:15}), we can find that the quench paths inside the ferromagnetic phase may cause the DQPTs. Fig.~\ref{fig:07} gives the Fisher zeros and the corresponding rate functions for the homogeneous chain ($w=0$) and two disorder chains with $w=0.01$ and $0.5$, respectively. In the case of homogeneous chain, we can see two intersections on each Fisher zeros line with the imaginary axis. The corresponding critical times are $t^{a}_{0}\approx2.58$ and $t^{b}_{0}\approx25.57$. As can be seen in Fig.~\ref{fig:07} (b) for $w=0.01$, the diagram of Fisher zeros is almost the same as that in the homogeneous chain. Meanwhile, from the insert graph of Fig.~\ref{fig:07} (d), we see that except the values of rate functions, the critical times of the disorder chain are similar to that in the homogeneous chain. Therefore, for the disorder chain, the small disorder will not influence the DQPTs.  As the disorder strength increaseing, the Fisher zeros near the larger critical times will shrink toward the real axis. Eventually, there will be a Fisher zeros region on the imaginary axis of the complex time plane. For example, for the disorder strength $w=0.5$ [see Fig.~\ref{fig:07} (c)], we can only see one region of Fisher zeros on the imaginary. This indicates that the large disorder will destroy the original DQPTs and cause a critical region of the system.

\subsection{quench in the paramagnetic phase.}

\begin{figure}[t]
  \centering
  \includegraphics[width=1.0\linewidth]{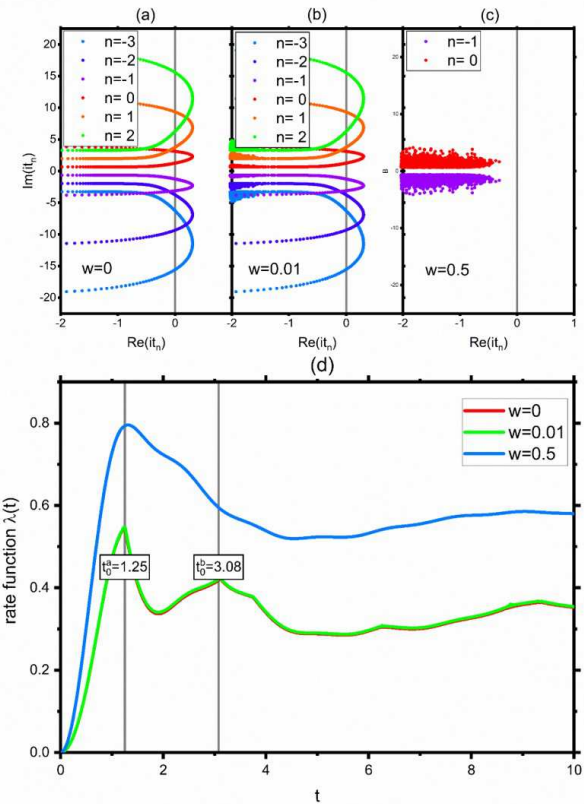}\\
  \caption{(a)-(c) The Fisher zeros of the homogeneous chian ($w=0$) and two disorder chains with $w=0.01$ and $0.5$ for the quench inside the paramagnetic phase from $(\gamma_{0}=0.89, h_{0}=1.1)$ to $(\gamma_{1}=-0.89, h_{1}=1.4)$. (d) The corresponding rate functions for different disorder strengths.}\label{fig:09}
\end{figure}

We can also find the DQPTs in the case of quench inside the paramagnetic phase. Fig.~\ref{fig:09} shows the Fisher zeros and the corresponding rate functions of the homogeneous chain ($w=0$) and two large disorder chains with $w=0.1$ and $0.5$, respectively. In the homogeneous, there two intersections on each Fisher zeros line with the imaginary axis. And the corresponding critical times are $t^{a}_{0}\approx1.25$ and $t^{b}_{0}\approx3.08$. Similar to that in the case of quench within the ferromagnetic phase, the small disorder can not affect the DQPTs. It can be verified from Fig.~\ref{fig:09} (a) and (b), where the Fisher zeros of $w=0.01$ are quite same as that of the homogeneous. Meanwhile, the rate function of $w=0.01$ is almost coincident with that of the homogeneous chain. As the disorder strength increasing, the Fisher zeros move to the left slowly, so that the Fisher zeros may not pass through the imaginary axis when the disorder strength is large enough. For example, there is no intersection between the Fisher zeros and the imaginary axis for $w=0.5$ from Fig.~\ref{fig:09} (c). This indicates that for the lager disorder systems, DQOTs does not occur.

\section{Summary}

In summary, we have discussed the influences of weak disorder on the DQPTs of five quench protocols in the transverse-filed anisotropic XY chain. We find that when the quench passes through the Ising transition, the small disorder will cause the appearance of new critical times. And as the disorder strength increasing, many critical times will occur to form a critical time interval, where the Loschmidt echo and Loschmidt amplitude will become non-analytic. When the quench crosses the anisotropic transition, the disorder will expand the larger critical times into the critical region. And the width of regions will increase by the disorder strength. When the quench passes through two kinds of quantum transition (Ising and anisotropic transitions), the small disorder will cause three additional critical times in the system. Similar to that in the case of quench across the Ising transition, as the disorder strength increasing, the new critical times will form regions. As to the quench within the ferromagnetic phase, the large disorder will turn the original two critical times into a critical region. And for the quench inside the paramagnetic phase, the large disorder will cause the DQPTs of the system to disappear.

It should be noticed that the results of our work are quite different from that in the Aubry-Andre model \cite{r34} and the Anderson model \cite{r35}. In their works, they only observe one critical time, which depends on the width of energy band. In addition, it is worth mentioning that the rate functions of the disorder system are always larger than that of the homogeneous system, which indicates that the disorder can accelerate the decay of Loschmidt echo with time.

\begin{acknowledgments}
The  work was supported by the National Natural Science Foundation of China (Grant Nos. 11575057 and  11975126).
\end{acknowledgments}


\begin{thebibliography}{111}

\bibitem{r1} W. H. Zurek, U. Dorner and P. Zoller, Phys. Rev. Lett. \textbf{95}, 105701 (2005).
\bibitem{r2} J. Dziarmaga, Adv. Phys. \textbf{59}, 1063 (2010).
\bibitem{r3} A. Polkovnikov, K. Sengupta, A. Silva and M. Vengalattore, Rev. Mod. Phys. \textbf{83}, 863 (2011).
\bibitem{r4} L. D'Alessio, Y. Kafri, A. Polkovnikov and M. Rigol, Adv. Phys. \textbf{65}, 239 (2016).
\bibitem{r5} M. Heyl,  Rep. Prog. Phys. \textbf{81} , 054001 (2018).
\bibitem{r6} D. A. Abanin, E. Altman, I. Bloch and M. Serbyn, Rev. Mod. Phys. \textbf{91}, 021001 (2019).
\bibitem{r7} M. Heyl, A. Polkovnikov, and S. Kehrein, Phys. Rev. Lett. \textbf{110}, 135704 (2013).
\bibitem{r8} S. Bhattacharjee and A. Dutta, Phys. Rev. B \textbf{97} (2018).
\bibitem{r9} U. Divakaran, S. Sharma, and A. Dutta,  Phys. Rev. E. \textbf{93}, 052133 (2016).
\bibitem{r10} S. Vajna and B. D\'{o}ra, Phys. Rev. B \textbf{89}, 161105 (2014).
\bibitem{r011} R. Jafari, Sci. Rep. \textbf{9}, 2871 (2019).
\bibitem{r11} B. Zunkovic, A. Silva and M. Fabrizio, Philos. Trans. R. Soc. A \textbf{374}, (2016).
\bibitem{r12} F. Andraschko and J. Sirker, Phys. Rev. B \textbf{89}, 125120 (2014).
\bibitem{r13} P. P. Mazza, J.-M. St\'{e}phan, E. Canovi, V. Alba, M. Brockmann and M. Haque, J. Stat. Mech. 013104 (2016).
\bibitem{r14} U. Marzolino and T. Prosen, Phys. Rev. B \textbf{96}, 104402 (2017).
\bibitem{r15} B. Pozsgay, J. Stat. Mech. p10028 (2013).
\bibitem{r16} C. Pineda, T. Prosen and E. Villase\"{n}or, New J. Phys. \textbf{16}, 123044 (2014).
\bibitem{r016} M. Schmitt and S. Kehrein, Phys. Rev. B \textbf{92}, 075114 (2015).
\bibitem{r17} U. Bhattacharya, S. Dasgupta and A. Dutta, Phys. Rev. E \textbf{90}, 022920 (2014).
\bibitem{r18} J. Klinder, H. Kessler, M. Wolke, L. Mathey and A. Hemmerich, PNAS \textbf{112}, 3290 (2015).
\bibitem{r024} J. M. Hickey, S. Genway and J. P. Garrahan, Phys. Rev. B \textbf{89}, 054301 (2014).
\bibitem{r19} S. Vajna and B. D\'{o}ra, Phys. Rev. B \textbf{91}, 155127 (2015).
\bibitem{r20} J. C. Budich and M. Heyl, Phys. Rev. B \textbf{93}, 085416 (2016).
\bibitem{r21} U. Bhattacharya and A. Dutta, Phys. Rev. B \textbf{96}, 014302 (2017).
\bibitem{r22} M. Heyl and J. C. Budich, Phys. Rev. B \textbf{96}, 180304 (2017).
\bibitem{r23} T. V. Zache, N. Mueller, J. T. Schneider, F. Jendrzejewski, J. Berges and P. Hauke, Phys. Rev. Lett. \textbf{122}, 050403 (2019).
\bibitem{r24} X. Peng, H. Zhou, B. B. Wei, J. Cui, J. Du and R. B. Liu, Phys. Rev. Lett. \textbf{114}, 010601 (2015).
\bibitem{r25} N. Fl\"{a}schner, D. Vogel, M. Tarnowski, B. S. Rem, D. S. L¨¹hmann, M. Heyl, J. C. Budich, L. Mathey, K. Sengstock, and C. Weitenberg,  Nat. Phys. \textbf{14}, 265 (2017).
\bibitem{r26} P. Jurcevic, H. Shen, P. Hauke, C. Maier, T. Brydges, C. Hempel, B. P. Lanyon, M. Heyl, R. Blatt, and C. F. Roos,  Phys. Rev. Lett. \textbf{119}, 080501 (2017).
\bibitem{r27} J. Zhang, G. Pagano, P. W. Hess, A. Kyprianidis, P. Becker, H. Kaplan, A. V. Gorshkov, Z. X. Gong, and C. Monroe,  Nature(London) \textbf{551}, 601 (2017).
\bibitem{r28} K. Wang, X. Qiu, L. Xiao, X. Zhan, Z. Bian, W. Yi, and P. Xue,  Phys. Rev. Lett. \textbf{122}, 020501 (2019).
\bibitem{r29} Y. Imry and S. Ma, Phys. Rev. Lett. \textbf{35}, 1399 (1975).
\bibitem{r30} A. Aharony, Y. Imry, and S. Ma,  Phys. Rev. Lett. \textbf{37}, 1364 (1976).
\bibitem{r31} M. Aizenman and J. Wehr,  Phys. Rev. Lett. \textbf{62}, 2503 (1989).
\bibitem{r32} A. Harris, J. Phys. C \textbf{7}, 3082 (1974).
\bibitem{r33} P. W. Anderson,  Phys. Rev. \textbf{109}, 1492 (1958).
\bibitem{r34} C. Yang, Y. Wang, P. Wang, X. Gao and S. Chen, Phys. Rev. B \textbf{95}, 184201 (2017).
\bibitem{r35} H. Yin, S. Chen, X. Gao and P. Wang, Phys. Rev. A \textbf{97}, 033624 (2018).
\bibitem{r39} E. Lieb, T. Schultz, and D. Mattis, Ann. Phys. \textbf{16}, 407 (1961).
\bibitem{r40} P. Pfetury, Ann. Phys. \textbf{57}, 79C90 (1970).
\bibitem{r41} M. Zhong, P. Tong, Phys. Rev. A \textbf{84}, 052105 (2011).
\bibitem{r38} C. Karrasch and D. Schuricht, Phys. Rev. B \textbf{87}, 195104 (2013).
\bibitem{r42} H. T. Quan, Z. Song, X. F. Liu, P. Zanardi, and C. P. Sun,  Phys. Rev. Lett. \textbf{96}, 140604 (2006).
\end{thebibliography}
\end{document}